\documentstyle[epsf,bibnorm]{lamuphys}

\begin{document}
\title{Spintronic Spin Accumulation and Thermodynamics}

\author{A.H. MacDonald}

\institute{Department of Physics, Indiana University, Bloomington,
         IN 47405 }

\maketitle

\begin{abstract}

The spin degree of freedom can play an essential role in determining the
electrical transport properties of spin-polarized electron systems in 
metals or semiconductors.
In this article, I address the dependence of spin-subsystem
chemical potentials on accumulated spin-densities. 
I discuss both approaches which can be used to measure this fundamental 
thermodynamic quantity and the
microscopic physics which determines its value in several different systems.

\end{abstract}

\section{Introduction}
\label{sec:introduction}

The role of the electronic spin degree of freedom in theories of the electrical
transport properties of paramagnetic metals is passive; usually it as appears
only as an afterthought---a factor of two to account for spin degeneracy. All
this changes profoundly in electronic systems with substantial
spin-polarization, either spontaneous or induced, particularly so if the system is
either electrically or magnetically inhomogeneous.  Recently, interest in the
role of the electronic spin has increased, in part because of the possibility of
fabricating technologically useful magnetoresistive sensors and other devices
based on spin-dependent transport effects, particularly giant magnetoresistance
\cite{gmr} and tunnel magnetoresistance \cite{tmr}. {\em Spintronics}
\cite{physicstoday}, the study of spin-dependent electronic transport effects in
systems containing metallic ferromagnets, is now a large and active area of
basic and applied physics. In this article, I discuss the possibility
of using transport experiments not to make devices, but instead to measure a
fundamental thermodynamic property of a
spin-polarized electron system, the dependence of spin subsystem chemical
potentials on accumulated spin-densities.
This quantity can be important in modeling some
spin-dependent transport effects.  I will discuss several examples  
where it also provides a new and useful test of our understanding of the
microscopic physics of a spin-polarized itinerant electron system.

In Section~\ref{sec:thermodynamic}, I derive a
formally exact expression for the dependence of chemical potentials 
on subsystem densities by defining a spin-dependent
thermodynamic-density-of-states {\em matrix}.  For bulk three-dimensional 
charged particle systems, the total density is fixed by electroneutrality requirements.
In this case only the dependence of up and down spin chemical potentials on 
the difference between up and down spin densities is of 
interest.  I show that these two 
quantities can be expressed in terms of the differential 
magnetic susceptibility, and a less familiar quantity, the 
derivative of chemical potential with respect to external 
field.  The focus of the article is this latter quantity, which I 
will refer to as the {\em inverse magnetic compressibility} .
In Section~\ref{sec:spintronic}, I
discuss two spintronics experiments which can be exploited to measure its value
in particular systems.  In Section~\ref{sec:microscopic}, I discuss the  
microscopic physics which determines its value in three distinct 
spin-polarized electron systems.  Section~\ref{sec:summary} contains
a brief summary.

\section{Thermodynamic Density-of-States Matrix}
\label{sec:thermodynamic}

Non-equilibrium spin accumulation \cite{oldspinaccum,johnson} due to electronic
transport occurs generically in inhomogeneous spin-polarized electron systems
and is a ubiquitous feature of spintronics. Any theory of spin accumulation
requires, explicitly or implicitly, a model for the relationship between the up
and down spin densities and their chemical potentials. Linearizing around the
equilibrium state, we can write 
\begin{eqnarray}
d \mu_{\uparrow} &=& (D_{\uparrow}^{-1}+ F_{\uparrow,\uparrow}) \: d
n_{\uparrow} + F_{\uparrow,\downarrow} \: d n_{\downarrow} \nonumber \\
d \mu_{\downarrow} &=& F_{\downarrow,\uparrow} \: d n_{\uparrow} +
(D_{\downarrow}^{-1}+ F_{\downarrow,\downarrow}) \: d n_{\downarrow}.
\label{tdosinv}
\end{eqnarray}
where $\mu_{\sigma}$ is the spin-$\sigma$ chemical potential and $n_{\sigma}$ is
the density of spin-$\sigma$ electrons. In these equations, I treat
$n_{\uparrow}$ and $n_{\downarrow}$ as separate thermodynamic variables,
something which is useful in discussing spin-accumulation since the processes
which establish equilibrium between spin-$\uparrow$ and spin-$\downarrow$
subsystems are often slow. The spin-quantization axis has been chosen to lie
along the direction of net spin-polarization \cite{collinearcaveat}. 

It is normally convenient to measure the local chemical potential of a charged
particle system from the local electrostatic potential and this common
convention is implicit throughout these notes. Accordingly
\begin{equation}
\mu_{\sigma} \equiv \frac{1}{V} \frac{\partial
F(T,n_{\uparrow},n_{\downarrow})}{\partial n_{\sigma}}
\label{chempotdef}
\end{equation}
with the free-energy per volume, $F/V$, calculated excluding any electrostatic
contributions. The matrix of coefficients in Eq.~(\ref{tdosinv}) is accordingly
given by the matrix of second derivatives of $F/V$ with respect to $n_{\sigma}$,
with the electrostatic term (which would diverge because of the long-range of
the electron-electron interaction) neglected. (Note that
$F_{\uparrow,\downarrow} = F_{\downarrow,\uparrow}$.) To make contact with
familiar descriptions of spin accumulation, I have introduced the
band-theory spin-dependent density-of-states per volume, $D_{\sigma}$. If
contributions due to correlation effects were neglected, the relationship
between chemical potentials and densities would be diagonal in spin indices and
only the density-of-states terms would appear on the right-hand-side of
Eq.~(\ref{tdosinv}).  The density-of-states contribution to the spin-$\sigma$ 
chemical potential change is simply the result of changing the filling
of the spin-$\sigma$ energy band.
I will refer to the additional correlation terms in
Eq.~(\ref{tdosinv}), $F_{\sigma,\sigma'}$, as local-field corrections. 
The fact that the local-field corrections are in
general off-diagonal in the spin indices can have a qualitative importance.
Eq.~(\ref{tdosinv}) is formally exact, however, the values
for the local field corrections, are not known in general, and their computation
is a challenge to theory. In the following Section, I discuss two experiments
which measure a particular combination of these coefficients.

Eq.~(\ref{tdosinv}) can be inverted to express spin-dependent density changes in
terms of spin-dependent chemical potential changes
\begin{equation}
d n_{\sigma} = \sum_{\sigma'} {\cal D}_{\sigma,\sigma'} d \mu_{\sigma'}
\label{tdosdef}
\end{equation}
where
\begin{eqnarray}
{\cal D}_{\sigma,\sigma} &=& \frac{ D_{\sigma} ( 1 + D_{\bar{\sigma}}
F_{\bar{\sigma},\bar{\sigma}})}{1 + D_{\sigma} F_{\sigma,\sigma} +
D_{\bar{\sigma}} F_{\bar{\sigma},\bar{\sigma}} - D_{\sigma} D_{\bar{\sigma}}
F_{\sigma,\bar{\sigma}}^{2}} \nonumber \\ {\cal D}_{\sigma,\bar{\sigma}} &=& -
\frac{ D_{\sigma} D_{\bar{\sigma}} F_{\bar{\sigma},\sigma}}{1 + D_{\sigma}
F_{\sigma,\sigma} + D_{\bar{\sigma}} F_{\bar{\sigma},\bar{\sigma}}- D_{\sigma}
D_{\bar{\sigma}} F_{\sigma,\bar \sigma}^{2}},
\label{tdos}
\end{eqnarray}
where $\bar{\sigma} = \downarrow$ if $\sigma = \uparrow$ and {\em vice-versa}.
${\cal D} \equiv \sum_{\sigma,\sigma'} {\cal D}_{\sigma,\sigma'}$, the rate of
change of total density with chemical potential when the spin-subsystems are in
equilibrium, is the total 
thermodynamic density-of-states of an electron system. 
If interactions are
neglected ${\cal D} = D_{\uparrow} + D_{\downarrow}$. In spintronics it is
useful to generalize this concept by defining a thermodynamic density-of-states
{\em matrix} as in Eq.~(\ref{tdos}).

To evaluate the inverse magnetic compressibility, I add to the Hamiltonian a Zeeman
coupling \cite{noorbital} term which 
contributes $ - g^{*} \mu_{B} H (n_{\uparrow} - n_{\downarrow})/2$ to the free
energy per unit volume.  Here $g^{*}$ is the system's g-factor, $\mu_{B}$ is the
electron Bohr magneton, and $H$ is the field strength. It is convenient to use a
notation where $\mu_{\sigma}$ is defined as the chemical potential
without its Zeeman contribution, whereas $\mu$ is the full field-dependent
chemical potential. Then the condition for equilibrium between up and down spins
is 
\begin{equation}
\mu = \mu_{\uparrow} - g^{*} \mu_{B} H/2 = \mu_{\downarrow} + g^{*} \mu_{B} H/2,
\label{equil}
\end{equation}
or differentiating with respect to field strength:
\begin{equation}
\frac{\partial \mu}{\partial g^{*} \mu_{B} H} = \frac{\partial
\mu_{\uparrow}}{\partial g^{*} \mu_{B} H} -\frac{1}{2} = \frac{\partial
\mu_{\downarrow}}{\partial g^{*} \mu_{B} H}+\frac{1}{2}
\label{derivs}
\end{equation}
At fixed total electron density ($d n_{\uparrow} +
d n_{\downarrow} = 0$), differentiating Eq.~(\ref{tdosdef})
with respect to $g^{*} \mu_{B} H$ gives
\begin{equation}
\frac{\partial \mu_{\uparrow}/\partial H}{\partial \mu_{\downarrow} / \partial
H} = - \frac{{\cal D}_{\uparrow,\downarrow} + {\cal
D}_{\downarrow,\downarrow}}{{\cal D}_{\uparrow,\uparrow} + {\cal
D}_{\downarrow,\uparrow}} = - \frac{D_{\uparrow}^{-1} +
F_{\uparrow,\uparrow}-F_{\uparrow,\downarrow}}{D_{\downarrow}^{-1} +
F_{\downarrow,\downarrow} - F_{\uparrow,\downarrow}}.
\label{ratio}
\end{equation}
Combining Eq.~(\ref{ratio}) and Eq.~(\ref{derivs}), I obtain that for 
fixed total density
\begin{equation}
\frac{\partial \mu}{\partial g^{*} \mu_{B} H} = \frac{D_{\uparrow}^{-1} -
D_{\downarrow}^{-1} +F_{\uparrow,\uparrow} -F_{\downarrow,\downarrow}} {2 [
D_{\uparrow}^{-1} + D_{\downarrow}^{-1} + F_{\uparrow,\uparrow} +
F_{\downarrow,\downarrow} - 2 F_{\uparrow,\downarrow}]}.
\label{dmudh}
\end{equation}
Similar considerations lead to the following expression for the differential
magnetic susceptibility,
\begin{equation}
\chi_{S} = \frac{g^{*}\mu_{B}}{2}\frac{\partial (n_{\uparrow} - n_{\downarrow})}{
\partial H} = \frac{(g^{*} \mu_{B})^{2}} { D_{\uparrow}^{-1} +
D_{\downarrow}^{-1} + F_{\uparrow,\uparrow} + F_{\downarrow,\downarrow} - 2
F_{\uparrow,\downarrow}}.
\label{dxidh}
\end{equation}
Both of these expressions are formally exact.
I show in the following paragraph that the dependence of 
the chemical potentials on spin accumulation is specified
by the inverse magnetic compressibility and $\chi_{S}$.

Theories of spin accumulation contain in general three elements: i) a theory
of the spin and space dependent transport coefficients which lead to non-equilibrium
spin-densities, ii) a theory for the disequilibration ($\mu_{\uparrow} -
\mu_{\downarrow}$) produced by these spin densities, and iii) a theory of the
relaxation process which attempts to establish equilibrium between the spin
subsystems. The thermodynamic property we are discussing is related to the
second element.  Assuming electroneutrality ($d n_{\downarrow}
= - d n_{\uparrow}$ ) it follows from Eq.~(\ref{tdosinv}) that 
\begin{eqnarray}
\frac{d \mu_{\uparrow}}{d (n_{\uparrow}-n_{\downarrow})} &=&
\frac{D_{\uparrow}^{-1} +
F_{\uparrow,\uparrow}-F_{\uparrow,\downarrow}}{2}\nonumber\\
\frac{d \mu_{\downarrow}}{d (n_{\uparrow}-n_{\downarrow})} &=& -
\frac{D_{\downarrow}^{-1} +
F_{\downarrow,\downarrow}-F_{\uparrow,\downarrow}}{2}.
\label{pdosdiff}
\end{eqnarray}
Note that $d(\mu_{\uparrow}-\mu_{\downarrow})/d(n_{\uparrow}-n_{\downarrow}) =
(g^{*} \mu_{B} )^{2}/ 2 \chi_{S} $. This relationship should not be a surprise
since an external magnetic field maintains a chemical potential difference
$\mu_{\uparrow}-\mu_{\downarrow}=g^{*} \mu_{B} H$ and induces a magnetization
per volume $m = \chi_{S} H = g^{*} \mu_{B} (n_{\uparrow}-n_{\downarrow})/2$. The
linear relationship between $\mu_{\uparrow} - \mu_{\downarrow}$ and the
non-equilibrium spin accumulations is thus completely characterized by $\chi_{S}$.
It is only if we want to know the chemical potential shifts of up-spin and
down-spin subsystems individually that the inverse magnetic compressibility 
is required.  The individual
chemical potential shifts driven by a non-equilibrium spin accumulation are:
\begin{eqnarray*}
\frac{4 \chi_{S}}{(g^{*} \mu_{B})^{2}} \frac{d \mu_{\uparrow}}{d
(n_{\uparrow} - n_{\downarrow})} &=& 1 + 2\frac{d \mu}{d (g^{*} \mu_{B} H)} =
\frac{D_{\uparrow}^{-1} + F_{\uparrow,\uparrow} -
F_{\uparrow,\downarrow}}{D_{\uparrow}^{-1} + D_{\downarrow}^{-1} +
F_{\uparrow,\uparrow} + F_{\downarrow,\downarrow} - 2 F_{\uparrow,\downarrow}}\\
\frac{4 \chi_{S}}{(g^{*} \mu_{B})^{2}} \frac{d \mu_{\downarrow}}{d
(n_{\uparrow}-n_{\downarrow})} &=& - 1 + 2\frac{d \mu}{d (g^{*} \mu_{B} H)} =
-\frac{D_{\downarrow}^{-1}+F_{\downarrow,\downarrow}
- F_{\uparrow,\downarrow}}{D_{\uparrow}^{-1}+D_{\downarrow}^{-1}
+ F_{\uparrow,\uparrow}+F_{\downarrow,\downarrow}-2F_{\uparrow,\downarrow}}
\end{eqnarray*}

\section{Spintronic Thermodynamic Measurements}
\label{sec:spintronic}

\subsection{Field-dependent Coulomb Blockade Peaks}

The first type of experiment I discuss was pioneered by Ono and co-workers
\cite{ferrosets} and takes advantage of the equally spaced conductance peaks
which occur in Coulomb blockade devices.  The
experimental geometry is that of a single-electron-transistor (SET) in which
current flow from source to drain through a small metallic particle is influenced
by a gate voltage.  In general source lead, drain lead, and metallic particle
can be either paramagnetic or ferromagnetic.  For definiteness, I assume
that the only the small metallic particle is ferromagnetic and address a situation
which is simpler than what has been encountered in experiments 
by also assuming a single domain.
The following paragraph provides a simplified explanation the operation of a  
SET which is sufficient for present purposes.

The dependence of the ground-state energy of an isolated metallic grain on its
net charge is dominated by an electrostatic contribution and has the form
\begin{equation}
E_{0}(N) =  \frac{e^{2} (N - N_{0})^{2}}{2C} + V \epsilon_{0} (N/V) 
\label{coulblock}
\end{equation}
where $e(N - N_{0})$ is the net charge on the grain and the effective capacitance
of the grain $C \sim R$ where $R$ is the grain diameter. Here $\epsilon_{0}(n)$
is the energy per unit volume $V$ calculated for a macroscopic grain which is
electrically neutral. The conductance for current flow through the grain between
lead electrodes is sharply peaked when the addition
energy of the island is equal to the chemical potential in the lead electrodes
$\mu_L$,
{\em i.e.} when
\begin{equation}
\mu(N) \equiv E_{0}(N + 1)-E_{0}(N) = \mu_{L}.
\label{peakcondition}
\end{equation}
In a SET the chemical potential
for electrons on the metallic grain is manipulated by a gate voltage $U$:
\begin{equation}
E_{0}(N) \to  \frac{e^{2} (N - N_{0})^{2}}{2C} + V \epsilon_{0} (N/V) - N e U. 
\label{chargeenergy}
\end{equation}
As $U$ is varied, the equilibrium number of particles on the grain changes.
The number of particles in the grain's ground state 
changes between $N$ and $N+1$ when Eq.(~\ref{peakcondition}) is satisfied:
\begin{equation}
e U^{*}_{N} = e^{2} (N+1/2)/C +  \mu(N/V) -\mu_{L}.
\label{cbpeak}
\end{equation}
A peak occurs in the source-drain conductance at this value of the gate 
voltage.  In this equation 
$\mu(n)$ is the bulk chemical potential of an electrically neutral system,
the quantity examined in Section~\ref{sec:thermodynamic}.

As discussed in Section~\ref{sec:thermodynamic}, $\mu(n)$ is   
field-dependent in general. It follows from Eq.~(\ref{cbpeak}) that 
provided the field-dependence of the lead chemical potentials can 
be ignored
\begin{equation}
\frac{d \mu(n)}{d (g^{*} \mu_{B} H)} = \frac{d (eU^{*}_{N})}{d( g^* \mu_B
H)};
\label{exptdmudh}
\end{equation}
the quantity of interest can simply be read off the gate voltage dependence of
the Coulomb blockade conductance peak.   More realistic models are complicated
by geometry dependent cross capacitances between different circuit elements
which invalidate the simple relationship between the chemical potential on the
ferromagnetic grain and the gate voltage assumed here. However these two
quantities are still proportional and the proportionality constant can be sorted
out experimentally by measuring the spacing between Coulomb blockade conductance
peaks at fixed voltage. Estimates of $d (\mu_{0}(n)/ d (g^{*} \mu_{B} H) $ have
already been obtained \cite{ferrosets} using this approach for one ferromagnetic
transition metal. 

\subsection{Field-dependent double-layer compressibility measurements}

The second potential experiment I discuss is a 
variant on one which has been used in the past
\cite{eisenstein,papadakis,patel,ying} to measure the compressibility of
two-dimensional electron gas layers. It exploits techniques which
have been developed \cite{eisenref} to make separate contact to nearby
two-dimensional electron layers. The experimental set up can be thought of as
a parallel plate capacitor, where one plate is a metal or heavily doped
semiconductor and the second
plate consists of two separately contacted two-dimensional electron layers, one
on top of the other and closer to the metallic gate.  A change in the charge
density on the surface of the metal induces an opposing charge density
distributed between the two two-dimensional electron layers. The equilibrium
condition which determines the distribution of charge is:
\begin{equation}
\mu_{T}(n_{T}) = \mu_{B}(n_{B}) + 4 \pi e^{2} d (n - n_{T} - n_{0}/2).
\label{equildl}
\end{equation}
Here $n_{T}$ is the areal density in the top two-dimensional layer, $n_{B}$ is
the density in the bottom two-dimensional layer, $d$ is the separation between
the two layers, and $n = n_{T} + n_{B}$ is the total density in the two-layers.
The second term on the right hand side of Eq.~(\ref{equildl}) is the electrostatic
potential drop due to the electric field which exists between top and bottom
two-dimensional layers. Note that the electric field between the gate and the
top layer has magnitude $4 \pi e (n - n_{0}/2) $, {\em i.e.} $e n$ equals the
surface charge density on the metallic gate up to a constant. If the top
two-dimensional layer is held at ground, the gate voltage $V_G$ is therefore
proportional to $n$. The magnitude of the electric field below the bottom
two-dimensional layer, $ 4 \pi e n_{0}/2$, does not change during the
experiment. 

To determine the compressibility \cite{eisenstein}, it is necessary only to 
measure the current which flows to the
bottom layer when the gate voltage $V_G$ (or equivalently $n$) is changed. 
Differentiating Eq.~(\ref{equildl}) with respect to $n$ with $n_{B}+n_{T}=n$ 
leads to \cite{eisenstein} a relationship between a fundamental thermodynamic quantities and a
conveniently measurable dimensionless experimental quantity exploited in 
previous experimental work:
\begin{equation}
\frac{d n_{B}}{d n} = \frac{d \mu_{T}/ d n_{T}}{4 \pi e^{2} d + d \mu_{T}/d
n_{T} + d \mu_{B}/d n_{B}}
\label{jimsexpt}
\end{equation}
Since the first term in the denominator of Eq.~(\ref{jimsexpt}) 
is normally dominant 
the experiment provides a direct measurement of the dependence of
chemical potential on density, inversely proportional to the compressibility,
for the top layer.

I propose using the same experimental setup to measure the dependence of
the charge in the bottom layer on magnetic field at fixed 
gate voltage.  Such a measurement can be 
related to the inverse magnetic compressibility;    
differentiating Eq.~(\ref{equildl}) with respect to field at fixed $n$ I find
that 
\begin{equation}
\frac{d n_{B}}{d (g^{*} \mu_{B} H) } = \frac{d \mu_{T}/ d (g^{*} \mu_{B} H) - d
\mu_{B}/ d (g^{*} \mu_{B} H)}{4 \pi e^{2} d + d \mu_{T}/d n_{T} + d \mu_{B}/d
n_{B}}.
\label{newexpt}
\end{equation}
Since the denominator is dominated by the first term, the experiment provides
a direct measurement of the difference between the rate of change
chemical potential with field in top and bottom 2D layers.  

\section{Microscopic Theory of the Thermodynamic Density-of-States Matrix}
\label{sec:microscopic}

\subsection{Band Ferromagnets}

In order to contextualize the issues raised by these relatively new measurement
possibilities, I first discuss the estimate for $d\mu/d(g^*\mu_BH)$ which follows from the 
simplest possible mean-field theory of a ferromagnetic metal, 
Stoner-Wohlfarth \cite{stoner} theory.  In modern work
this approach is wrapped in the cloak of spin-density-functional
theory \cite{dft}.  In the Stoner-Wohlfarth theory, quasiparticle
spin-up and spin-down energies for each band and for each wavevector in the 
crystal's Brillouin-zone are split
by an amount
\begin{equation}
\Delta = 2 \mu_{B} H_{eff} = 2 ( I m  + \mu_{B} H).
\label{splitting}
\end{equation}
The effective magnetic field includes an exchange contribution which is
proportional to the magnetic moment per Bohr magneton per volume of the system,
\begin{equation}
m  = \frac{M}{V \mu_{B}}= n_{\uparrow}-n_{\downarrow}.
\label{magmom}
\end{equation}
where $M = \mu_{B} (N_{\uparrow} - N_{\downarrow})$ is the total magnetic moment.
The exchange integral $I$ is a phenomenological material property which
is characteristic of a given system.  In a given field, $m$ and the chemical
potential $\mu$ are determined by self-consistently occupying spin-split
bands.  At $T=0$ the self-consistent mean-field equations are
\begin{eqnarray}
m &=&  \int_{\mu-\Delta/2}^{\mu + \Delta/2} d \epsilon \: D(\epsilon) \nonumber\\
n &=& 2 \int_{- \infty}^{\mu - \Delta/2} d \epsilon D(\epsilon) + \int_{\mu -
\Delta/2}^{\mu + \Delta/2} d \epsilon D(\epsilon)
\label{meanfield}
\end{eqnarray}
where $D(\epsilon)$ is the
density-of-states per volume per spin at $\Delta = 0$. 
Let $\Delta_{0}$ and
$I$ denote the self-consistent exchange splitting of the band and the
self-consistent value of $m$ in the absence of an external field respectively.
Expanding $\mu$ and $\Delta$ to first order in $H$, I find that
\begin{equation}
\frac{d \mu}{ d (g^{*} \mu_{B} H)} = \frac{D_{\downarrow}^{-1} -
D_{\uparrow}^{-1}}{2 [ D_{\downarrow}^{-1} + D_{\uparrow}^{-1} -4 I] }.
\label{derivative}
\end{equation}
Here $D_{\uparrow} = D(\mu_{0} + \Delta_{0}/2)$ and $ D_{\downarrow} = D(\mu_{0}
- \Delta_{0}/2)$ are the majority and minority spin densities of states in the
absence of a field.  

This expression can be recognized as a special case of the general result
derived in Section~\ref{sec:thermodynamic} which is obtained by holding the
bands rigid and taking $F_{\uparrow,\downarrow} = - F_{\uparrow,\uparrow} = -
F_{\downarrow,\downarrow} = I$.  In the formally exact theory, the three
local-field-factors are independent.  Density-functional theory \cite{dft} 
is a practical approach to many-particle physics which has been applied 
successfully to evaluate the thermodynamic properties of metals and 
underpins the modern theory \cite{bandtheory} of ferromagnetic transition metals.
In density-functional theory, the local field factors arise from
an interplay between band-structure details and exchange-correlation
single-particle potentials.  Expectations based on the electron gas 
case, discussed for two-dimensions below, suggest that the diagonal
local-field-factors should be negative and larger in magnitude than
the positive off-diagonal local-field-factor.  It is not at all 
obvious that these expectations apply to transition metals, especially
because the majority spins at the Fermi energy tend to have 
predominantly itinerant $s$-electron character while the minority
spins tend to have predominantly localized $d$-electron character.
Comparison of theoretical and experimental values for 
both the differential magnetic susceptibility, and the magnetic 
compressibility, presents a serious and interesting challenge to 
the density-functional theory of metallic magnetism.

\subsection{Zero Field Two-Dimensional Electron Gas}

The ground state of the two-dimensional electron gas is not ferromagnetic except
at extremely low densities \cite{tanatar,sergio}.  Spin-polarization can, however, be
induced by application of an external magnetic field \cite{orbital}. I now 
discuss a simple theory for the value of $d\mu/d(g^{*}\mu_{B} H)$ in a spin-polarized 
two-dimensional electron gas which is based on the
Hartree-Fock approximation \cite{stern} for its energy:
two-dimensional electron system:
\begin{equation}
\frac{E_{n_{\uparrow},n_{\downarrow}}}{A} =
\frac{n_{\uparrow}^{2}+n_{\downarrow}^{2}}{2 D_{2D}}  - \frac{8 e^{2}
[n_{\uparrow}^{3/2} +n_{\downarrow}^{3/2}]} {3 \pi^{1/2}}
\label{2deng}
\end{equation}
Here $D_{0} = m^{*}/(2 \pi \hbar^{2})$ is the density-of-states per area $A$ in
a 2D system.  The first term on the right hand side of Eq.(~\ref{2deng})
is the band energy and the second term is the exchange energy.  Differentiating
twice with respect to density we obtain for the diagonal local-field-correction
\begin{equation}
F_{\sigma,\sigma} = - \frac{2 e^{2} n_{\sigma}^{-1/2}}{\pi^{1/2}}.
\label{2dlocfield}
\end{equation}
The off diagonal local-field-corrections vanishes in this approximation because
of the neglect, in the Hartree-Fock approximation,
of correlations between electrons of opposite spin.  In the
presence of an external magnetic field the partitioning of density between
spin-subsystems is determined by setting $\mu_{\uparrow} - \mu_{\downarrow} =
g^{*} \mu_{B} H $ where $\mu_{\sigma} = n_{\sigma}/D_{2D} - 4 e^{2}
n_{\sigma}^{1/2}/\pi^{1/2}$.  For electron-gas density\cite{densparam} parameter $r_{s}=2$,
the Hartree-Fock ground state is
paramagnetic and the spin-polarization is initially linear in field. At stronger
fields, the exchange term causes the minority-spin chemical potential to
increase with decreasing density.  
The polarization must then increases more
rapidly with field in order to establish the required chemical potential
difference.  The polarization eventually jumps from a partial value to 
full polarization when $g^{*} \mu_{B} H \sim
0.034 {\rm Ry}$.  (The atomic energy unit 
$1 {\rm Ry} = e^{4} m^{*} / 2 \epsilon^2 \hbar^{2}$ is $\approx 5.5 {\rm meV}$ in
GaAs.) 

\begin{figure}
\centerline{\epsfxsize=2in
     \epsffile{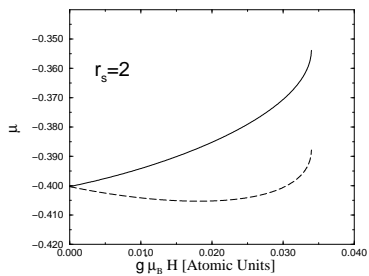}}
\caption{Hartree-Fock approximation majority-spin (solid line) and minority-spin
(dashed line) chemical potentials {\em vs.} Zeeman coupling strength for a
two-dimensional electron gas with density parameter $r_{s}=2$.}
\label{cpvszeeman}
\end{figure}

\begin{figure}
\centerline{\epsfxsize=2in
   \epsffile{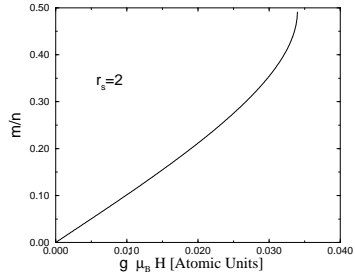}}
\caption{Hartree-Fock approximation $m/n \equiv (n_{\uparrow} -
n_{\downarrow})/(n_{\uparrow} + n_{\downarrow})$ for a two-dimensional electron
gas with density parameter $r_{s} = 2$ {\em vs.} Zeeman coupling strength $g^{*}
\mu_{B} H$ in atomic units ($e^{4} m^{*} / 2 \hbar^{2}$). For GaAs the Zeeman
coupling strength in atomic (Ry) units is $0.0052 H$[Tesla].  The system jumps to 
full polarization at $g^* \mu_B H \sim 0.034 {\rm Ry}$.} 
\label{polvszeeman}
\end{figure}

The results for the Zeeman coupling denpendence of subsystem chemical potentials
and spin-polarization, shown in Fig.(~\ref{cpvszeeman}) and Fig.(~\ref{polvszeeman})
respectively, can lead to substantial values of 
$d\mu/d (g^{*} \mu_{B} H)$ which can even exceeds one 
over a wide range of Zeeman coupling strength. 
This is despite the fact that the minority and
majority spin density-of-states are identical in two-dimensions so that 
the inverse magnetic compressibility would be identically zero for a non-interacting
electron system.
\begin{figure}
\centerline{\epsfxsize=2in
     \epsffile{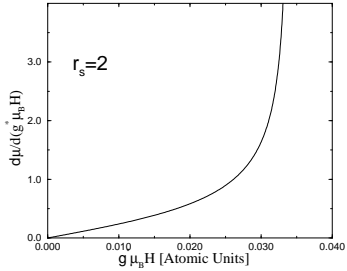}}
\caption{Hartree-Fock approximation $d\mu/d(g^{*} \mu_{B} H)$ {\em vs.} Zeeman
coupling strength, for a two-dimensional electron gas with density parameter
$r_{s}=2$.  For $g^{*} \mu_{B} H > 0.034$, only the majority spins are occupied 
and $d\mu /d(g^{*} \mu_{B} H) \equiv -0.5$.}
\label{dcpdhvszeeman}
\end{figure}
Numerical results for the inverse magnetic compressibility for 
this illustrative example are shown in Fig.(~\ref{dcpdhvszeeman}).
Of course, we do not expect the 
Hartree-Fock approximation theory to be reliable in an electron gas for $r_s > 1$.
The large spin-polarizations at relatively weak Zeeman 
couplings in Fig.~(\ref{polvszeeman}),
occur because the Hartree-Fock approximation ground state is ferromagnetic
for $r_s > \pi/2^{1/2}$. We know from reliable quantum Monte Carlo
calculations\cite{tanatar,sergio} that, in reality, this instability does not
occur until much lower densities are reached. Nevertheless illustrative
calculation does suggest what should be expected in low-density two-dimensional
electron gas layers. For modern samples, which have high mobility at densities a
few $\times 10^{10} {\rm cm}^{-2}$, it is clear that a substantial degree of
spin-polarization (and possibly complete spin polarization) can be achieved in
laboratory superconducting magnets. Measurement of the inverse magnetic
compressibility, coupled with simultaneous measurements of the spin
polarization, will provide a strict test of our understanding of magnetism in
strongly correlated electron gas systems.

\subsection{Quantum Hall Ferromagnets}

In the quantum Hall regime, the 2D electron density is measured in units of the
density which can be accommodated by a full Landau level; the Landau level
filling factor $\nu \equiv (2 \pi \ell^{2}) n$ where $\ell \equiv (\hbar c/eB)^{1/2}$ 
is the magnetic length.  At Landau level filling factor
$\nu=1$, the 2D electron system has a strong ferromagnet ground state, {\it i.e.}
{\em all} spins
can be aligned with the field by arbitrarily weak Zeeman coupling.
Quantum Hall ferromagnets have unusual charged
excitations known as Skyrmions \cite{skyrmtheory,skyrmexpt}. Skrymions are topologically
non-trivial configurations of the magnetization orientation distribution which occur
in any 2D magnet, but carry an electrical charge $e$ only in the case of 
quantum Hall ferromagnets.  Skyrmions have an internal \cite{microskyrm} integer
quantum number $K$ which specifies the number reversed spins in their interior.
When the Landau level filling factor is close to $\nu = 1$, the low energy states of
the system can be described in terms of Skyrmion degrees of freedom.

A simple model of a 2DES, valid at low temperature in this regime because Skyrmions are 
dilute, is obtained by ignoring Skyrmion-Skyrmion interactions.
The following grand-canonical ensemble expressions specify the occupation probabilities of the
$N_{\phi} = A/ (2 \pi \ell^{2})$ Skyrmion quasielectron and quasihole states
with $K$ excess reversed spins:
\begin{eqnarray}
n_{Ke} &=& f(\epsilon_{K} + K \mu_{\uparrow} - (K+1) \mu_{\downarrow} ) \nonumber \\
n_{Kh} &=& f(\epsilon_{K} + (K+1) \mu_{\uparrow} - K \mu_{\downarrow}).
\label{skyrmionoccs}
\end{eqnarray}
Here $f(\epsilon) = (\exp(\epsilon/k_{B}T)+1)^{-1}$ is a Fermi factor
\cite{caveat}, $\epsilon_{K}$ is the energy of a Skyrmion quasiparticle, $(2 \pi
\ell^{2})^{-1} $ is the density of a full Landau level, and we have chosen the
zero of energy so that quasielectron and quasihole skyrmion states have the same
energy\cite{kunyang}. When the spin subsystems are in equilibrium, 
we can use Eqs.~(\ref{skyrmionoccs}) to
calculate the chemical potential, given the Landau level filling factor. The
Landau level filling factor is increased by quasielectron excitations and
decreased by quasihole excitations:
\begin{equation}
\nu = 1 + \sum_{K} (n_{Ke} - n_{Kh}).
\label{nuskyrm}
\end{equation}
Eq.(~\ref{skyrmionoccs}) follows from the property that formation of the $K-th$
quasielectron Skyrmion requires the addition of $K+1$ spin-down electrons and
the removal of $K$ spin-up electrons from the $\nu=1$ ground state, while
formation of the $K-th$ quasihole Skyrmion requires the addition of $K$ spin-down
electrons and the removal of $K+1$ spin-up electrons. For non-interacting
electrons only the $K=0$ quasiparticles occur; for typical 2DES's, on the other
hand, the lowest energy quasiparticles 
have $K=3$; these quasiparticles dominate the low-temperature properties of 
the system for $\nu$ close to $1$.  From Eqs.(~\ref{skyrmionoccs}) I obtain the
following thermodynamic density-of-states matrix:, 
\begin{eqnarray}
{\cal D}_{\uparrow,\uparrow} &=& (2 \pi \ell^{2})^{-1} \sum_{K} \left[(K+1)^{2}
\Delta(\epsilon_{K} + \mu) +K^{2} \Delta(\epsilon_{K} -\mu)\right]\\
{\cal D}_{\uparrow,\downarrow} &=& - (2 \pi \ell^{2})^{-1} \sum_{K} K (K+1)
\left[\Delta(\epsilon_{K}+\mu) + \epsilon_{K} - \mu)\right]
\label{thermodos}
\end{eqnarray}
where $\Delta(x) = {\rm sech}^{2}(x/2)/4 k_{B} T$.  

\begin{figure}
\centerline{\epsfxsize=2in
     \epsffile{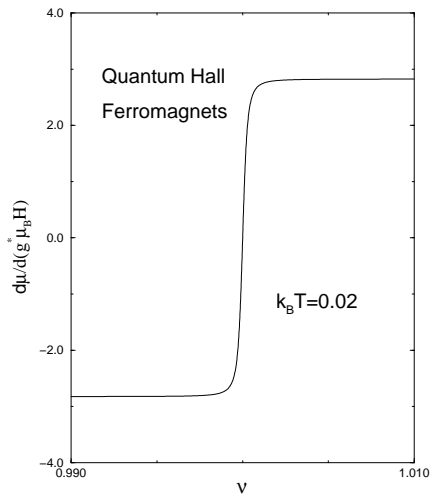}}
\caption{$d \mu/ d(g^{*} \mu_{B} H)$ {\em vs.} Landau level filling factor for a
quantum Hall system near $\nu=1$. at temperature $k_B T =0.02 e^2/\ell$.}
\label{qhall}
\end{figure}

These expressions have been used previously to analyze
\cite{qhspinbottleneck} spin bottlenecks which have been observed
\cite{ashoori} in the quantum Hall regime. Here I evaluate the
dependence of the equilibrium chemical potential on Zeeman coupling strength.
The possibility of measuring this quantity in two-dimensional electron
systems was discussed in the previous section.
Zeeman coupling adds \cite{skyrmtheory,kunyang} to the
Skyrmion quasiparticle energies, $\epsilon_{K} \to 
\epsilon_{K} + g^{*} \mu_{B} H (K+1/2)$.  In order for the filling factor
to be held fixed as the Zeeman coupling strength varies, the chemical potential
must change:
\begin{equation}
\frac{\partial \mu}{\partial (g^{*} \mu_{B} H)} = - \frac{\partial n/\partial
(g^* \mu_B  H)|_{\mu}} {\partial n/\partial \mu|_{H}} = \frac{\sum_{K} (K+1/2)
[\Delta(\epsilon_{K}-\mu) - \Delta(\epsilon_{K}+\mu)]} {\sum_{K}
[\Delta(\epsilon_{K}-\mu) + \Delta(\epsilon_{K}+\mu)]}
\label{dmudhskyrm}
\end{equation}
We see in Fig.~(\ref{qhall})
that for an ideal disorder-free two-dimensional electron system, $d \mu/d(g^{*}
\mu_{B} H)$ is typically larger than one and changes sign when the filling
factor crosses $\nu=1$.  These results were calculated using Skrymion state
energies evaluated by Palacios {\it et. al.} \cite{palacios}.
The value of this quantity gives information about the
Skyrmion system complementary to that available from other
experiments\cite{skyrmexpt}. Its value depends both on the average $K$ value of
the Skrymions present in the system, the quantity revealed by Knight shift
measurements, but also on the way in which the distribution of $K$ values
changes with field. Unlike transport experiments, this thermodynamics
measurement would be sensitive to Skrymions localized by disorder in the system.

\section{Summary}
\label{sec:summary}

In this article, I have discussed the dependence of chemical potential 
on external magnetic field in a spin-polarized electron system.  
This quantity appears directly in theories of electron 
systems disturbed by non-equilibrium spin accumulations generated by 
spin and space-dependent electronic transport.  I have discussed 
techniques which can be used to measure this fundamental 
thermodynamic quantity, both in two-dimensional
electron systems and in ferromagnetic metals.  I believe that 
its measurement can provide an important and interesting test
of theory in several quite different spin-polarized 
electron systems.


\begin{thebibliography}{99}
\bibitem{collinearcaveat} I will assume here that we are interested only in
states with a collinear spin-polarization configuration.
\bibitem{noorbital} In these notes I will always neglect the coupling of an
external field to orbital degrees of freedom of the electrons. Although this
coupling does not play an essential role in the physics discussed here, the role
it plays in particular experiments should always be carefully assessed.
\bibitem{densparam}  In 2D dimensions the electron gas density is 
related to the dimensionless parameter $r_s$ by
$ n=1/(\pi r_s^2 a_0^2)$ where $a_0 \equiv \hbar^2 \epsilon /(m^* e^2)$ is
the host system dielectric constant.  In GaAs $a_0 \approx 10 {\rm nm}$.
\bibitem{orbital} In two-dimensions it is possible to largely the coupling to
orbital degrees-of-freedom of an external magnetic field oriented in the plane
of the 2D system can usually be neglected.
\bibitem{caveat}  Our primary interest is in the dilute 
high-temperature regime where the use of Fermi quantum statistics for 
skyrmions does not play any essential role.
\end{thebibliography}
\end{document}